\documentclass[12pt]{iopart}

\usepackage{iopams}
\usepackage{setstack}
\usepackage{graphicx}
\usepackage{pict2e}
\usepackage{color}
\usepackage{hyperref}

\newcommand{\Li}{\mathrm{Li}}
\renewcommand{\Re}{\mathrm{Re}}
\newcommand{\half}{\tfrac{1}{2}}

\newcommand{\affiliation}[1]{\address{#1}}
\renewcommand{\pacs}[1]{\noindent\textbf{PACS numbers:} #1}
\newcommand{\keywords}[1]{\noindent\textbf{Keywords:} #1}
\newcommand{\tfrac}[2]{\mbox{\small$\frac{#1}{#2}$}}
\renewcommand{\text}[1]{\mathrm{#1}}

\begin{document}

\title[Current fluctuations for TASEP on the relaxation scale]{Current fluctuations for totally asymmetric exclusion on the relaxation scale}
\author{Sylvain Prolhac}
\affiliation{Laboratoire de Physique Th\'eorique; IRSAMC; UPS; Universit\'e de Toulouse; France\\Laboratoire de Physique Th\'eorique; UMR 5152; Toulouse; CNRS; France}

\begin{abstract}
The fluctuations of the current for the one-dimensional totally asymmetric exclusion process with $L$ sites are studied in the relaxation regime of times $T\sim L^{3/2}$. Using Bethe ansatz for the periodic system with an evolution conditioned on special initial and final states, the Fourier transform of the probability distribution of the fluctuations is calculated exactly in the thermodynamic limit $L\to\infty$ with finite density of particles. It is found to be equal to a sum over discrete realizations of a scalar field in a linear potential with coupling constant equal to the rescaled time $T/L^{3/2}$.\\\\
\\\\
\end{abstract}

\maketitle

At equilibrium, computing expectation values of physical observables requires the evaluation of a partition sum over the set $\Omega$ of all micro-states, whose number is often exponential in the system size. This scheme is valid even for systems with only a few degrees of freedom, where it has the well defined meaning of having the system in contact with a thermal bath. In the thermodynamic limit, it becomes even more general due to a kind of universality called equivalence of ensembles, that states that the details of the coupling between the system and the heat bath become irrelevant for large systems with short range effective interactions.

The situation is more complicated for systems out of equilibrium. In the context of Markov processes with finite number of states, the stationary probabilities in the non-equilibrium steady state require in general a summation over spanning trees on $\Omega$ \cite{S1976.1}, whose number typically grows exponentially with the number of micro-states; it is only for very special exactly solvable models that this huge summation simplifies to "only" an exponentially large sum, such as in the case of models with a matrix product representation \cite{BE2007.1} for the stationary state.

At large scales, however, when the number of degrees of freedom goes to infinity, some kind of universality is expected as in the equilibrium case. A prominent example is one-dimensional driven diffusive systems \cite{SZ1998.1} and growth models \cite{HHZ1995.1} in the Kardar-Parisi-Zhang (KPZ) universality class \cite{SS2010.4,KK2010.1,C2011.1}, for which many exact results have been obtained in the last twenty years for fluctuations of current and interface height. Two regimes have received much attention: the stationary regime, attained in the long time limit, for which large deviations of the fluctuations have been characterized \cite{DL1998.1,P2010.1,GLMV2012.1}, and the transient regime of infinite volume, where probability distributions of the fluctuations have been obtained \cite{J2000.1,TW2009.1,D2010.1,CLDR2010.1,SS2010.3,ACQ2011.1} and were later observed experimentally in growth of turbulent phases in liquid crystals \cite{TS2010.1,TSSS2011.1}.

The crossover between these two regimes is characterized by the system size $L$ and the observation time $T$ going both to infinity with $T\sim L^{3/2}$. This corresponds to the scale of relaxation to stationarity, with the dynamical exponent $3/2$ of one-dimensional KPZ universality. We compute the large scale limit of the fluctuations of the current in a specific model, the totally asymmetric simple exclusion process (TASEP) \cite{D1998.1,GM2006.1}, conditioned on special initial and final states. Our main result (\ref{Gtx asymptotics}) is very similar to an equilibrium expectation value, with a sum over discrete realizations of a scalar field $\varphi$ in a linear potential.

We consider TASEP defined on a periodic one-dimensional lattice with $L$ sites, on which $N$ undistinguishable classical particles are placed. The exclusion constraint requires that each site has at most one particle. The dynamics consists of random hops of particles from any site $i$ to the site $i+1$ at its right. In a small time interval $dt$, each particle attempts to move with probability $dt$ provided it has no neighbour on its right: moves breaking the exclusion constraint are forbidden. The total number of particles in the system is conserved, and we call $\rho=N/L$ the density of particles. In the following, we are interested in the thermodynamic limit $L,N\to\infty$ with $\rho$ kept fixed.

For technical reasons, it is convenient to condition the evolution both on the initial and the final state. We choose them respectively equal to $\mathcal{C}_{X}$ and $\mathcal{C}_{Y}$, where $\mathcal{C}_{i}$ is the configuration where sites $i,i+1,\ldots,i+N-1$ are occupied, the other sites being empty. The numbers $X$ and $Y$ are not required to be between $1$ and $L$ at the moment, although the evolution only depends on $Y-X$ modulo $L$. The initial and final states correspond to initial and final density profiles equal to $1$ in an interval and $0$ in the rest of the system. This choice has no influence in the stationary situation $T\gg L^{3/2}$ \cite{DL1998.1}; in the transient regime $T\ll L^{3/2}$, however, several universality subclasses exist \cite{C2011.1}.

We consider the total time-integrated current $Q$ through the system, equal to the total number of hops of particles anywhere in the system between time 0 and time $T$. This observable is related to the current $Q_{\text{b}}$ through a bond between two sites, say $(L,1)$, by $Q=LQ_{b}+\sum_{j=1}^{N}(x_{j}-x_{j}^{0})$, where $x_{j}^{0}$ and $x_{j}$, taken between $1$ and $L$, are the positions of the particles respectively in the initial and the final configuration. At large times, $Q$ grows typically proportionally to $T$, with a coefficient $\rho(1-\rho)L$ equal to the sum over all sites $i$ of the probability that site $i$ is occupied while site $i+1$ is empty. We are interested in the fluctuations around this mean value, since they give some indication on the universal process emerging at large scales.

The total current $Q$ depends on the whole evolution of the system between time $0$ and time $T$, and not only on the state of the system at time $T$. Its evolution in time follows from the master equation for the probability $P_{T}(\mathcal{C},Q)$ that the system is at time $T$ in the configuration $\mathcal{C}$ with a value $Q$ for the current. It is useful \cite{DL1998.1} to consider the quantities $F_{T}(\mathcal{C},\gamma)=\sum_{Q=-\infty}^{\infty}\rme^{\gamma Q}P_{T}(\mathcal{C},Q)$ with parameter $\gamma$ conjugate to the current. The probability $P_{T}(\mathcal{C})$ that the system is in configuration $\mathcal{C}$ at time $T$ is equal to $F_{T}(\mathcal{C},0)$. Writing $F_{T}(\mathcal{C},\gamma)$ in terms of the conditional probability $P_{T}(Q|\mathcal{C})=P_{T}(\mathcal{C},Q)/P_{T}(\mathcal{C})$, we observe that the generating function $G(\gamma)=F_{T}(\mathcal{C}_{Y},\gamma)/F_{T}(\mathcal{C}_{Y},0)$ is the average of $\rme^{\gamma Q}$ over all evolutions conditioned on ending in configuration $\mathcal{C}_{Y}$ at time $T$ and starting in configuration $\mathcal{C}_{X}$ at time $0$.

Since TASEP dynamics is Markovian, the transition rates between configurations are independent of the whole evolution, and in particular of $Q$. It implies that $F_{T}(\mathcal{C},\gamma)$ also evolves in time by a linear master equation. Writing the $F_{T}(\mathcal{C},\gamma)$ in a vector $|F_{T}(\gamma)\rangle=\sum_{\mathcal{C}}F_{T}(\mathcal{C},\gamma)|\mathcal{C}\rangle$, the evolution in time of $|F_{T}(\gamma)\rangle$ is given by $\partial_{T}|F_{T}(\gamma)\rangle=M(\gamma)|F_{T}(\gamma)\rangle$, and the generating function can be computed from the matrix $M(\gamma)$ as
\begin{equation}
\label{GF[M]}
G(\gamma)=\big\langle\rme^{\gamma Q}\big\rangle=
\frac{\langle\mathcal{C}_{Y}|\rme^{TM(\gamma)}|\mathcal{C}_{X}\rangle}
{\langle\mathcal{C}_{Y}|\rme^{TM(0)}|\mathcal{C}_{X}\rangle}\;.
\end{equation}

The generating function can be evaluated further by inserting into (\ref{GF[M]}) a decomposition of the identity operator on the eigenstates of $M(\gamma)$ and $M(0)$. We write $E_{r}(\gamma)$, $\langle\psi_{r}(\gamma)|$, $|\psi_{r}(\gamma)\rangle$ respectively for the eigenvalues and corresponding left and right (unnormalized) eigenvectors of $M(\gamma)$. Since $M(\gamma)$ is not Hermitian, the eigenvalues are complex numbers, and the left and right eigenvectors are not related in a simple way by transposition. One has
\begin{equation}
\label{Y*eTM*X}
\langle\mathcal{C}_{Y}|\rme^{TM(\gamma)}|\mathcal{C}_{X}\rangle=
\sum_{r}\rme^{TE_{r}(\gamma)}\frac{\langle\mathcal{C}_{Y}|\psi_{r}(\gamma)\rangle\langle\psi_{r}(\gamma)|\mathcal{C}_{X}\rangle}{\langle\psi_{r}(\gamma)|\psi_{r}(\gamma)\rangle}\;.
\end{equation}
The dynamics of TASEP is known to be integrable in the sense of quantum integrability, also called stochastic integrability \cite{S2012.1} in this context. Both eigenvalues and eigenvectors of $M(\gamma)$ can be obtained \cite{DL1998.1} using the Bethe ansatz technique. Each eigenstate $r$ is characterized by $N$ complex numbers, the Bethe roots $y_{j}$, $j=1,\ldots,N$, which are required to satisfy the $N$ Bethe equations $\rme^{L\gamma}(1-y_{j})^{L}=(-1)^{N-1}\prod_{k=1}^{N}(y_{j}/y_{k})$. Then, the corresponding eigenvalue of $M(\gamma)$ and of the translation operator are equal to
\begin{equation}
\label{EP[y]}
E_{r}(\gamma)=\sum_{j=1}^{N}\frac{y_{j}}{1-y_{j}}
\quad\text{and}\quad
\rme^{\frac{2\rmi\pi p_{r}}{L}}=\rme^{N\gamma}\prod_{j=1}^{N}(1-y_{j})\;,
\end{equation}
while the scalar products in (\ref{Y*eTM*X}) are equal to \cite{P2014.3}
\begin{eqnarray}
\label{psi*psi}
&& \frac{\langle\mathcal{C}_{Y}|\psi(\gamma)\rangle\langle\psi(\gamma)|\mathcal{C}_{X}\rangle}{\langle\psi(\gamma)|\psi(\gamma)\rangle}
=(-1)^{\frac{N(N-1)}{2}}\rme^{\frac{2\rmi\pi p_{r}(Y-X+1-N)}{L}}\\
&&\hspace{43mm} \times\frac{(\rme^{N\gamma}\prod_{j=1}^{N}y_{j}^{-1})^{N-1}\prod_{1\leq j<k\leq N}(y_{j}-y_{k})^{2}}{\frac{L}{N}\bigg(\sum_{j=1}^{N}\frac{y_{j}}{N+(L-N)y_{j}}\bigg)\prod_{j=1}^{N}\Big(L-N+\frac{N}{y_{j}}\Big)}\;.\nonumber
\end{eqnarray}
The latter formula follows from the Gaudin determinant \cite{GMCW1981.1,K1982.1} for scalar products of Bethe states. It has a particularly simple form for TASEP, where the usual very complicated determinant essentially reduces to the denominator in the right hand side of (\ref{psi*psi}).

The Bethe equations of TASEP can be solved using the following procedure: let us define the quantity $b=\gamma+\frac{1}{L}\,\sum_{j=1}^{N}\log y_{j}$. Then, there exists wave numbers $k_{j}$, integers (half-integers) if $N$ is odd (even) such that $y_{j}=g^{-1}(\exp(-b+2\rmi\pi k_{j}/L))$, where $g^{-1}$ is the inverse of the function $g:y\mapsto(1-y)/y^{\rho}$ \cite{P2014.1}. The wave numbers are required to be distinct modulo $L$ because of the exclusion constraint. The expression above for the $y_{j}$ is very convenient to study the thermodynamic limit: it allows to write both the eigenvalue (\ref{EP[y]}) and the scalar product (\ref{psi*psi}) in terms of sums over $j$ of functions of $k_{j}/L$, on which the Euler-Maclaurin formula can be used to derive large $L$ asymptotic expansions.

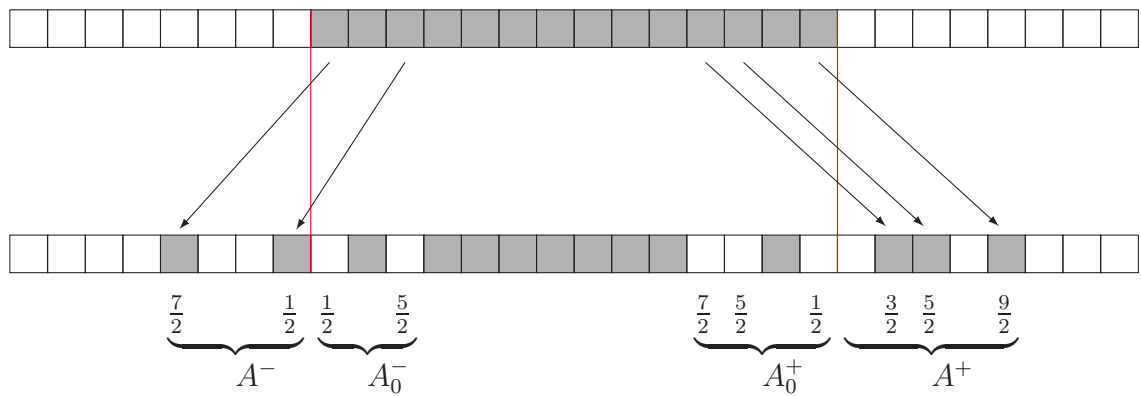
\begin{figure}
  \begin{center}
    \begin{picture}(150,46)
      \put(0,40){\color[rgb]{0.7,0.7,0.7}\polygon*(40,0)(110,0)(110,5)(40,5)}
      \multiput(0,40)(5,0){30}{\polygon(0,0)(5,0)(5,5)(0,5)}
      \put(20,10){\color[rgb]{0.7,0.7,0.7}\polygon*(0,0)(5,0)(5,5)(0,5)}
      \put(35,10){\color[rgb]{0.7,0.7,0.7}\polygon*(0,0)(5,0)(5,5)(0,5)}
      \put(45,10){\color[rgb]{0.7,0.7,0.7}\polygon*(0,0)(5,0)(5,5)(0,5)}
      \put(0,10){\color[rgb]{0.7,0.7,0.7}\polygon*(55,0)(90,0)(90,5)(55,5)}
      \put(100,10){\color[rgb]{0.7,0.7,0.7}\polygon*(0,0)(5,0)(5,5)(0,5)}
      \put(115,10){\color[rgb]{0.7,0.7,0.7}\polygon*(0,0)(5,0)(5,5)(0,5)}
      \put(120,10){\color[rgb]{0.7,0.7,0.7}\polygon*(0,0)(5,0)(5,5)(0,5)}
      \put(130,10){\color[rgb]{0.7,0.7,0.7}\polygon*(0,0)(5,0)(5,5)(0,5)}
      \multiput(0,10)(5,0){30}{\polygon(0,0)(5,0)(5,5)(0,5)}
      \put(40,10){\color[rgb]{1,0,0}\line(0,1){35}}
      \put(110,10){\color[rgb]{1,0,0}\line(0,1){35}}
      \put(42.5,38){\vector(-0.9,-1){20}}
      \put(52.5,38){\vector(-0.65,-1){14.5}}
      \put(92.5,38){\vector(1.1,-1){24}}
      \put(97.5,38){\vector(1.1,-1){24}}
      \put(107.5,38){\vector(1.1,-1){24}}
      \put(21,3.5){$\frac{7}{2}$}
      \put(36,3.5){$\frac{1}{2}$}
      \put(41,3.5){$\frac{1}{2}$}
      \put(51,3.5){$\frac{5}{2}$}
      \put(91,3.5){$\frac{7}{2}$}
      \put(96,3.5){$\frac{5}{2}$}
      \put(106,3.5){$\frac{1}{2}$}
      \put(116,3.5){$\frac{3}{2}$}
      \put(121,3.5){$\frac{5}{2}$}
      \put(131,3.5){$\frac{9}{2}$}
      \put(21,1.5){$\underbrace{\hspace{18\unitlength}}$}
      \put(41,1.5){$\underbrace{\hspace{13\unitlength}}$}
      \put(91,1.5){$\underbrace{\hspace{18\unitlength}}$}
      \put(111,1.5){$\underbrace{\hspace{23\unitlength}}$}
      \put(30,-5){$A^{-}$}
      \put(47.5,-5){$A_{0}^{-}$}
      \put(100,-5){$A_{0}^{+}$}
      \put(122.5,-5){$A^{+}$}
    \end{picture}
  \end{center}
  \caption{Characterization of the first excited states by $4$ sets of positive half integers $A_{0}^{+}$, $A_{0}^{-}$, $A^{+}$, $A^{-}$ verifying the constraint $|A_{0}^{+}|=|A^{+}|$, $|A_{0}^{-}|=|A^{-}|$.}
  \label{fig choice k}
\end{figure}

Since we are interested in the crossover regime $T\sim L^{3/2}$, we only need to consider the eigenstates with eigenvalues having a real part scaling as $L^{-3/2}$ in the thermodynamic limit with fixed density of particles. We call \textit{first eigenstates} these eigenstates, which are infinitely many. The first one is the stationary state. It corresponds to the choice of consecutive wave numbers $k_{j}=k_{j}^{0}$ with $k_{j}^{0}=j-(N+1)/2$, which is exactly like a filled Fermi sea of spinless fermions in one dimension. It was shown in \cite{P2014.1} that all other first eigenstates correspond to independent excitations at finite distance of both ends of the Fermi sea. More precisely, the first excited states are built by removing from $\{k_{j}^{0},j=1,\ldots,N\}$ a finite number of $k_{j}$'s at a finite distance of $\pm N/2$ and adding the same number of $k_{j}$'s at a finite distance of $\pm N/2$ but outside of the interval $[-N/2,N/2]$. On each side, the number of $k_{j}$'s removed and added must be equal. They can be described by $4$ finite sets of positive half integers $A_{0}^{\pm},A^{\pm}\subset\mathbb{N}+\half$: the $k_{j}$'s removed from $\{k_{j}^{0},j=1,\ldots,N\}$ are the elements of $N/2-A_{0}^{+}$ and $-N/2+A_{0}^{-}$, while the $k_{j}$'s added are the elements of $N/2+A^{+}$ and $-N/2-A^{-}$, see figure \ref{fig choice k}. The cardinals of the sets verify the constraints $m_{r}^{+}=|A_{0}^{+}|=|A^{+}|$, $m_{r}^{-}=|A_{0}^{-}|=|A^{-}|$. We define $m_{r}=m_{r}^{+}+m_{r}^{-}$. In the following, we use the symbol $r$ to label the first eigenstates, as a shorthand for $(A_{0}^{+},A^{+},A_{0}^{-},A^{-})$.

To each elementary excitation represented by a half-integer $a$, we associate quantities $\chi_{a}(u)=\rmi\sqrt{2u+4\rmi\pi a}$ and $\overline{\chi}_{a}(u)=-\rmi\sqrt{2u-4\rmi\pi a}$. The large $L$ asymptotics of the eigenvalue (\ref{EP[y]}) and of the scalar product (\ref{psi*psi}) can be expressed nicely \cite{P2014.1,P2014.3} in terms of the field $\varphi_{r}$ equal to
\begin{eqnarray}
\label{phi[A,zeta]}
&& \varphi_{r}(u)=\sum_{a\in-\mathbb{N}-\half}\!\!\chi_{a}(u)+\sum_{a\in A_{0}^{+}}\chi_{a}(u)+\sum_{a\in A^{-}}\chi_{a}(u)\nonumber\\
&&\hspace{10mm} +\sum_{a\in-\mathbb{N}-\half}\!\!\overline{\chi}_{a}(u)+\sum_{a\in A_{0}^{-}}\overline{\chi}_{a}(u)+\sum_{a\in A^{+}}\overline{\chi}_{a}(u)\;.
\end{eqnarray}
The divergent infinite sums are made sense of by the zeta regularization $\pm\rmi\sum_{a\in-\mathbb{N}-1/2}\sqrt{u\pm2\rmi\pi a}=\sqrt{2\pi}\rme^{\pm\rmi\pi/4}\zeta(-\half,\half\pm\tfrac{\rmi u}{2\pi})$, which comes directly from the Euler-Maclaurin formula. The Hurwitz zeta function is the analytic continuation for $\nu\neq1$ of $\zeta(\nu,z)=\sum_{j=0}^{\infty}(j+z)^{-\nu}$. The functions $\varphi_{r}$ have branch points $\pm\rmi\pi$ coming from the $\zeta$ functions, as well as branch points in $\pm2\rmi\pi(\mathbb{N}+\half)$ from the square roots. The branch cuts of $\varphi_{r}$ can always be taken equal to $(-\rmi\infty,-\rmi\pi]\cup[\rmi\pi,\rmi\infty)$.

The derivations of the asymptotics involving (\ref{phi[A,zeta]}) use Euler-Maclaurin formula applied to functions with logarithmic and square root singularities coming from the function $g^{-1}$ in terms of which the Bethe equations are solved. A generalization with square roots of Stirling's formula for the $\Gamma$ function is used crucially in order to handle some of these singularities. One has
\begin{eqnarray}
\label{EM sqrt(Stirling)}
&& \sum_{j=1}^{M}\log(\sqrt{j+d}\pm\sqrt{v})
\simeq \frac{M\log M}{2}-\frac{M}{2}\pm2\sqrt{v}\sqrt{M}\\
&&\hspace{37mm} +\frac{1}{2}\log\frac{\sqrt{2\pi}M^{d-v+\half}}{\Gamma(d-v+1)}
\pm\int_{0}^{v}\rmd u\,\frac{\zeta(\half,u+d-v+1)}{2\sqrt{u}}\;.\nonumber
\end{eqnarray}

With the relaxation scale in mind, we define a rescaled conjugate variable $s$ and a rescaled time $t$. We write
\begin{equation}
\gamma=\frac{s}{\sqrt{\rho(1-\rho)}L^{3/2}}
\qquad\text{and}\qquad
T=\frac{t\,L^{3/2}}{\sqrt{\rho(1-\rho)}}\;.
\end{equation}
Then, imposing that $\rme^{-\gamma JLT}G(\gamma)$ has a finite limit when $L\to\infty$ gives natural scalings for the distances and the mean value of the current $J$. One finds
\begin{equation}
Y-X=\frac{(1-2\rho)\,t\,L^{3/2}}{\sqrt{\rho(1-\rho)}}+\Big(x-\frac{1-2\rho}{2}\Big)L\;.
\end{equation}
and
\begin{equation}
J=\rho(1-\rho)-\frac{(\rho(1-\rho))^{3/2}}{t\sqrt{L}}\;.
\end{equation}
The large $L$ limit of the typical individual velocities of the particles $v_{\text{p}}=J/\rho$ is equal to $1-\rho$, which is positive. On the other hand, the scaling for the distances corresponds to a group velocity $v_{\text{g}}=(Y-X)/T$ asymptotically equal to $1-2\rho$, which can be either positive or negative. The sub-leading term introduces a reduced position $x$, defined modulo $1$ since the positions $X$ and $Y$ are defined modulo $L$.

The fluctuations $\xi_{t,x}$ of the current per site $Q/L$ are defined by subtracting the typical value and rescaling by the magnitude of the fluctuations, of order $T^{1/3}\sim\sqrt{L}$ in 1-dimensional KPZ universality. We define
\begin{equation}
\xi_{t,x}=\frac{Q/L-JT}{\sqrt{\rho(1-\rho)}\sqrt{L}}\;.
\end{equation}

The generating function of the fluctuations in the large $L$ limit is $G_{t,x}(s)=\lim_{L\to\infty}\langle\rme^{s\xi_{t,x}}\rangle=\lim_{L\to\infty}\rme^{-\gamma JLT}G(\gamma)$. From the asymptotic expansion of the eigenvalue \cite{P2014.1} (easily extended to non-zero $\gamma$ scaling as $L^{-3/2}$) and of the scalar product \cite{P2014.3}, one finds
\begin{equation}
\label{Gtx asymptotics}
\boxed{G_{t,x}(s)=\frac{1}{Z}\sum_{r}\frac{\omega_{r}\,\rme^{2\rmi\pi p_{r}x}}{\varphi_{r}'(\varphi_{r}^{-1}(s))}\,\rme^{-S_{s,t}[\varphi_{r}]}}\;.
\end{equation}
The summation is over all admissible choices for the sets $A_{0}^{\pm}$, $A^{\pm}$ that characterize the first eigenstates. The normalization constant $Z$ is such that $G_{t,x}(0)=1$. The functional $S_{s,t}$ is very similar to the action of a scalar field in a linear potential: $S_{s,t}[\varphi_{r}]=\lim_{\Lambda\to\infty}S_{s,t}^{\Lambda}[\varphi_{r}]-D_{t}^{\Lambda}$ with
\begin{equation}
\label{S[phi]}
S_{s,t}^{\Lambda}[\varphi_{r}]=-\int_{-\Lambda}^{\varphi_{r}^{-1}(s)}\!\!\!\rmd u\,\Big((\varphi_{r}'(u))^{2}+t\,\varphi_{r}(u)+1\Big)\;.
\end{equation}
The path of integration is required to avoid the branch cuts of $\varphi_{r}$: it has to cross the imaginary axis between $-\rmi\pi$ and $\rmi\pi$ when $\Re\,\varphi_{r}^{-1}(s)>0$. The integral in (\ref{S[phi]}) is divergent at $u\to-\infty$; the calculation of the asymptotics with Euler-Maclaurin formula provides the above regularization, where the divergent part is
\begin{equation}
D_{t}^{\Lambda}=\frac{4\sqrt{2}m_{r}t}{3}\Lambda^{3/2}-\Lambda-2\sqrt{2}\rmi\pi q_{r}t\sqrt{\Lambda}-2m_{r}^{2}\log\Lambda\;,
\end{equation}
with $q_{r}=\sum_{a\in A_{0}^{+}}a+\sum_{a\in A^{-}}a-\sum_{a\in A_{0}^{-}}a-\sum_{a\in A^{+}}a$. The total momentum $p_{r}$ in (\ref{Gtx asymptotics}) is equal for the first eigenstates to $p_{r}=\sum_{a\in A_{0}^{+}}a+\sum_{a\in A^{+}}a-\sum_{a\in A_{0}^{-}}a-\sum_{a\in A^{-}}a$. The combinatorial factor $\omega_{r}$ is given by
\begin{equation}
\omega_{r}=\frac{\omega(A_{0}^{+})\omega(A_{0}^{-})\omega(A^{+})\omega(A^{-})\omega(A_{0}^{+},A_{0}^{-})\omega(A^{+},A^{-})}{(-1)^{p_{r}+m_{r}}(\pi^{2}/4)^{-m_{r}^{2}}(2\pi)^{2m_{r}}}\;,
\end{equation}
with
\begin{equation}
\omega(A)=\prod_{a,a'\in A\atop a<a'}(a-a')^{2}\;,
\qquad
\omega(A,A')=\prod_{a\in A}\prod_{a'\in A'}(a+a')^{2}\;.
\end{equation}
The function $\varphi_{r}^{-1}$ is the inverse function of the field, $\varphi_{r}(\varphi_{r}^{-1}(s))=s$. It appears from Bethe ansatz since $\varphi_{r}^{-1}(s)$ is the large $L$ limit of $L(b-\rho\log\rho-(1-\rho)\log(1-\rho))$. The denominator $\varphi_{r}'(\varphi_{r}^{-1}(s))$ comes from the denominator in (\ref{psi*psi}). The linear potential in the action is a part of the eigenvalue $E_{r}(\gamma)$. Finally, the integral of $(\varphi_{r}')^{2}$ in the action follows from the asymptotics of the Vandermonde determinant of the Bethe roots in (\ref{psi*psi}), which uses (\ref{EM sqrt(Stirling)}) and a related formula with summation on two indices.

The probability distribution $P_{\xi}$ of the random variable $\xi_{t,x}$ in the large $L$ limit can be obtained from the generating function at imaginary argument by inverse Fourier transform $P_{\xi}(w)=\int_{-\infty}^{\infty}\frac{\rmd s}{2\pi}\,\rme^{\rmi sw}G_{t,x}(-\rmi s)$. Interestingly, making the change of variables $d=\varphi_{r}^{-1}(-\rmi s)$ in the integral over $s$, the Jacobian cancels the factor $\varphi_{r}'(\varphi_{r}^{-1}(-\rmi s))$ in (\ref{Gtx asymptotics}).

\begin{figure}
  \begin{center}
    \begin{tabular}{cc}
      \includegraphics[width=70mm]{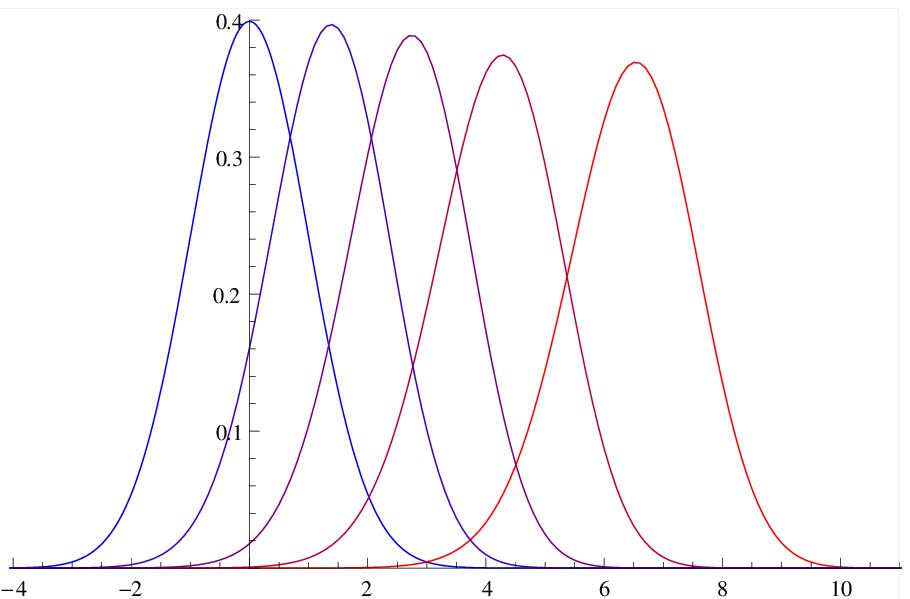}
      \begin{picture}(0,0)\put(-70,35){(a)}\end{picture}
      &
      \includegraphics[width=70mm]{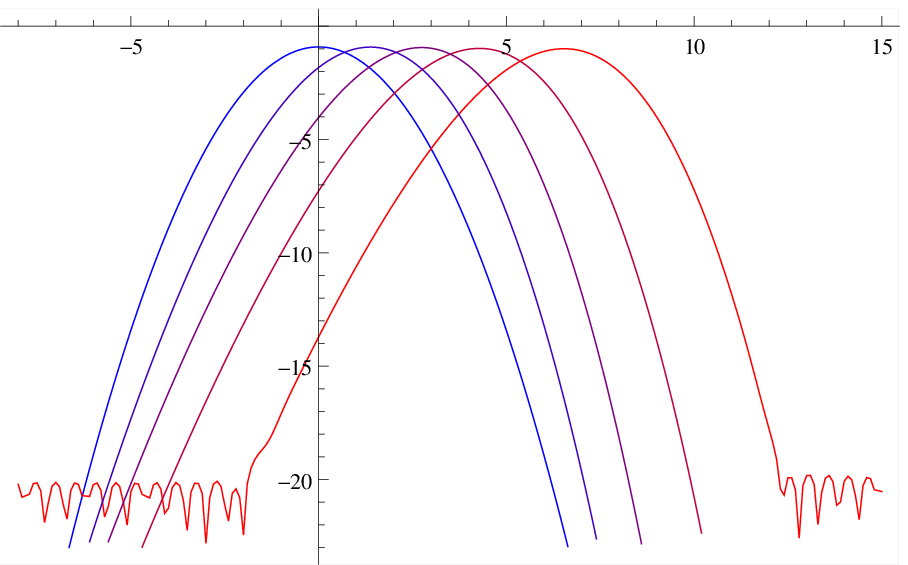}
      \begin{picture}(0,0)\put(-70,35){(b)}\end{picture}
    \end{tabular}
  \end{center}
  \caption{Probability distribution function (a) and its logarithm (b) for the current fluctuations $\sqrt{2}\,\pi^{-1/4}(\xi_{t,0}-t)/\sqrt{t}$, obtained by summing over the first $74$ eigenstates. The various curves correspond from right to left to rescaled times $t$ equal to $0.1$, $0.2$, $0.5$, $2$, and $+\infty$ where the distribution is a standard Gaussian. The artifacts at $t=0.1$ are caused by imperfect cancellations in the numerical evaluation of the Fourier transform.}
  \label{fig PDF}
\end{figure}

The short time behaviour of (\ref{Gtx asymptotics}) seems to involve significant contributions from many eigenstates, with subtle cancellations between them. We observe in figure \ref{fig PDF} that the probability distribution $P_{\xi}$ becomes very asymmetric in that limit. For large $t$ on the other hand, only the stationary state denoted by the index $r=0$ contributes to (\ref{Gtx asymptotics}), up to exponentially small terms in $t$. This eigenstate has $m_{0}=p_{0}=q_{0}=0$, $\omega_{0}=1$ and $\varphi_{0}(u)=-(2\pi)^{-1/2}\Li_{3/2}(-\rme^{u})$ using the relation between Hurwitz zeta function and polylogarithms, defined by $\Li_{\nu}(z)=\sum_{k=1}^{\infty}z^{k}/k^{\nu}$. The probability distribution of the random variable $\sqrt{2}\pi^{-1/4}(\xi_{t,x}-t)/\sqrt{t}$ converges to that of a standard Gaussian, see figure \ref{fig PDF}. In particular, its mean value approaches $0$ as $\sqrt{2}\pi^{1/4}/\sqrt{t}$ while the finite time correction to the variance is of order $1/t$. Beyond the Gaussian fluctuations, the large deviation function of Derrida and Lebowitz \cite{DL1998.1} corresponding to $\xi_{t,x}$ varying on the scale $t$ is recovered: the large $t$ limit of $t^{-1}\log\langle\rme^{s\xi_{t,x}}\rangle$ is equal to $G[\sqrt{2\pi}s]/\sqrt{2\pi}$, with $G$ defined by equations (20) and (21) in \cite{DL1998.1}.

\textit{Conclusions.} The fluctuations of the current have been computed for TASEP, a far from equilibrium model with interacting particles, on the temporal scale on which the relaxation to the stationary state occurs. From the Bethe ansatz calculation, a field $\varphi$ appears as a result of logarithmic and square root singularities in the Euler-Maclaurin formula. The generating function of the fluctuations of the current is expressed as a discrete path integral over realizations of $\varphi$, the action being that of a scalar field in a linear potential with coupling constant equal to time. Although a clear physical interpretation of the field $\varphi$ is currently missing, it seems appealing that fluctuations in a non-equilibrium model relaxing to its non-equilibrium steady state can be described at large scales in an equilibrium-like fashion, with a time-dependent potential maintaining the system out of equilibrium.

\vspace{10mm}
%\bibliographystyle{unsrt}
%\bibliography{/users/prolhac/bib/references.bib}

\end{document}